\documentstyle[prd,aps,12pt]{revtex}
\begin{document}
\draft
\title{\bf Lensing Effect of a Cosmic String in Chern-Simons Gravity}
\author{ E. Stedile and R. Duarte}
\address{ Department of Physics - UFPR, P. O. Box 19081 - 81531/990 Curitiba
PR Brazil}
\date{July 1996}
\maketitle
\def\ball#1{\mathaccent '27#1}
\vskip0.1truecm
\begin{abstract}
It is pointed out that any conformally transformed of a flat space-time metric
$\ball g_{ij} = f(x)\;\eta_{ij}$ is a solution to Witten's equation of
Chern-Simons gravity, which holds outside matter in (2+1) dimensions. It is
also shown that a simultaneous exterior solution to both Witten's and
Einstein's equations, yields the lensing effect of an isolated cosmic string,
if $f(x)$ is reduced to an arbitrary dimensionless constant. However, the
solution to Witten's equation with $f(x)$ being an arbitrary and continuous
function of space-time coordinates, also leads to an open circular path for a
light ray near the string.

\end{abstract}
\vskip0.1truecm

\vskip0.2truecm
Chern-Simons gravity (CSG) is a (2+1)-dimensional gauge model for gravity
\cite{Wit,Hor}, which is in essence different from general relativity (GR).
Such an approach has been formulated as a field theory in a three-dimensional
curved space-time, and it contrasts with GR which states that in three
space-time dimensions space-time is flat outside matter. Indeed, GR in this
case has some characteristic features that distinguish it from the more
familiar (3+1)-dimensional gravity \cite{Star}, since in (2+1)-dimensions GR
field equations lead to a flat space-time outside matter with a conical
symmetry, where Einstein's theory accomplishes for topological defects.

Otherwise, it has been shown \cite{Ach} that (2+1)-dimensional gravity is
equivalent to a gauge theory, with a pure Chern-Simons action, and a gauge
group $ISO(2,1), SO(3,1)$, or $SO(2,2)$ respectively, depending on the value
of the cosmological constant. We recall that when a cosmological constant is
included in (2+1)-dimensional GR, Minkowski space-time is replaced by either a
de Sitter space or an anti-de Sitter space, and $ISO(2,1)$ is replaced by
$SO(3,1)$ or $SO(2,2)$, respectively. In the present paper we derive possible
physical solutions to the field equation of Chern-Simons gravity, within a
particle dynamics framework, and afterwards we study the trajectory of photons
in the neighbourhood of an isolated cosmic string.

As has been pointed out \cite{Hor}, a gauge model for (2+1)-dimensional
gravity for the $SO(3,2)$ group has a Chern-Simons interpretation, where the
fundamental variable is the vielbein $e^a_i$ and the spin connection $w^a_i$
is a function of $e^a_i$, by requiring that $D_i e^a_j - D_j e^a_i = 0, ( i,
j, k$ are world indices and $a, b, c, $ are Lorentz indices). The topological
Chern-Simons action in such a model is constructed from the Riemann curvature
tensor, which results to be
\begin{equation} I = \int_{\cal M} \epsilon^{ijk} \left[ \omega_{ia} \bigl(
\partial_j \omega^a_k - \partial_k \omega^a_j \bigr) +
{2\over 3}\epsilon_{abc}\omega^a_i \omega^b_j \omega^c_k \right]\end{equation}
If we vary Eq.(1) with respect to $e^a_i$, we obtain the field equation
outside matter
\begin{equation} D_k W_{ij} - D_j W_{ik} = 0\end{equation}
where $W_{ij} = R_{ij} - (1/4) g_{ij} R$.
Notice that Eq.(2) is a conformally invariant equation, since its left-hand
side is the three-dimensional analogue of the Weyl tensor. Indeed, the
vanishing of Eq.(2) is the condition asserting that a (2+1)-dimensional
space-time outside matter is a conformally flat manifold in the CSG model.

Recall that the conformal group is the group of diffeomorphisms of
compactified Minkowski space that leaves the space-time metric invariant up to
a Weyl rescaling $\ball g_{ij}\rightarrow f(x)\;\eta_{ij}$, where $f(x)$ is
an arbitrary and continuous function of space-time coordinates.
In (2+1)-dimensions the conformal group has ten generators and it is
isomorphic to SO(3,2), in the case of a Lorentzian signature $(-, +, +)$.
Since Eq.(2) states that (2+1)-dimensional space-time is conformally flat,
then any conformally flat space-time metric given in the form $ds^2 =
f(t, r, \theta)\; (- c^2 dt^2 + dr^2 + r^2 d\theta^2)$ should be a solution to
that equation. Notice that the metric tensor has in this case only the
diagonal components non-vanishing {\it i. e.} $\ball g_{00} = - \ball g_{11} =
- f(t, r, \theta),\;\ball g_{22} = f(t, r, \theta)\; r^2$, and an easy
computation shows that $ D^i\ball g_{ij}= 0$, which means that $\omega^a_i$ is
a metric connection. Moreover, the $W_{ij}$ thus obtained satisfy Eq.(2). The
above metric form can be written as
\begin{equation} ds^2 = - f(t, r, \theta)\; \gamma^{-2}\; c^2\; dt^2
\end{equation}
where $\gamma = ( 1 - \beta^2)^{-1/2} = (1 - v^2/c^2)^{-1/2}$ and $v$ is the
velocity of a particle in this field, as defined in terms of the coordinate
time $t$.

With $ds^2$ given in Eq.(3) the relativistic action for a particle with rest
mass $m$ under the influence of gravity is $ S = i m c\int ds =\int L dt$,
where $L = - m c^2 \sqrt{f(t, r, \theta)( 1 - \beta^2)} $ is the Lagrangian
of this particle. The above Lagrangian turns into the Lagrangian of special
relativity for a free particle in a vanishing field approximation
$f(t, r, \theta)\rightarrow 1$.

Now we derive the relativistic particle dynamics within the present framework,
and afterwards we consider the motion of this particle in the neighbourhood
of an isolated cosmic string. Taking into account the Lagrangian of the
particle already obtained, we see that the spatial components of its
relativistic momentum are
\begin{equation} p^\alpha = {\partial L\over \partial v_\alpha } = \sqrt{f(x)}
\;\gamma \; m\; v^\alpha,\qquad (\alpha = 1, 2)\end{equation}
where $v^\alpha$ is the corresponding component of the particle's velocity.
The above result suggests that gravity emerges in the CSG framework through a
non-minimal coupling.

Under a Legendre transformation of $L$, the relativistic Hamiltonian of the
particle is
\begin{equation}  H = \sum^{2}_{\alpha=1} p_\alpha v^\alpha- L = \sqrt{f(x)}
\;\gamma\; m\; c^2\end{equation}
which represents the energy $E$ of the particle properly, in a conservative
case. Indeed, this is the relativistic quantity that is conserved during the
motion of the particle under the influence of gravity in the CSG approach.

In polar coordinates the particle's Lagrangian reads
\begin{equation} L = - \sqrt{f(x)}\; \biggl( 1- {\dot r^2 \over c^2} -
{r^2\dot\theta^2 \over c^2} \biggr)^{1\over 2 }\; m\; c^2\end{equation}
where the dot denotes derivative with respect to the time coordinate. The
canonical momenta are then
\begin{equation}  P_r = {\partial L\over\partial\dot r } = \sqrt{f(x)}\;\gamma
\; m\; \dot r, \qquad P_\theta = {\partial L\over\partial\dot\theta } =
\sqrt{f(x)}\;\gamma\; N = \Lambda\end{equation}
Here $\Lambda$ plays the role of a relativistic angular momentum of the
particle in the presence of gravity and $N = m r^2 \dot \theta $ is its
Newtonian angular momentum. If we write Euler-Lagrange equations with the
Lagrangian (6), we conclude that the canonical momentum $\Lambda $ is a
constant of motion. Notice that, although the Newtonian angular momentum is
still conserved in this case, the quantity we must deal with is $\Lambda$,
instead of $N$.

With the above results the Hamiltonian of the particle becomes
\begin{equation} H = P_r \dot r + P_\theta \dot\theta - L =
{c^2 P^2_r \over E} + {c^2 \Lambda^2\over E r^2} + {m^2\; c^4 f(x)\over E}
\end{equation}
where we have considered the energy given in Eq.(5). Since the action and
the Hamiltonian are related by $ H + \partial S / \partial t = 0$ for a
conservative case, then we can write $S$ in the form
\begin{equation} S = - E\; t + F (r,\Lambda,E ) + \Lambda\theta\end{equation}
where $F$ is an unknown function of $r$ and of the constants of motion.
Otherwise, since $ P_r = \partial S / \partial r = d F / d r$ and $ P_\theta =
\partial S / \partial \theta = \Lambda$, then Eq.(8) turns into
\begin{equation}{c^2\over E}\biggl({d F \over d r }\biggr)^2 +
{c^2 \Lambda^2\over E r^2} + {m^2 c^4 f(x)\over E} - E = 0 \end{equation}

Now, if we integrate Eq.(10) we obtain for the action (9)
\begin{equation}  S = \int \sqrt {{E^2\over c^2} - f(x) m^2 c^2 -
\Lambda^2/r^2} \;\; dr + \Lambda\theta  - E t\end{equation}

The path of a particle is derived from the condition $\partial S/\partial
\Lambda = 0$, which yields
\begin{equation} \theta = \int {\Lambda\; dr \over r^2
\sqrt{E^2/c^2 - f(x) m^2 c^2 - \Lambda^2 /r^2 }}\end{equation}
however, from Eqs.(4) and (5) we easily conclude that
\begin{equation} E = \sqrt{ p^2\; c^2 + f(x) m^2 c^4}\end{equation}
which states that for photons {\it i. e.} $ m = 0$ we still have $E = p\; c$
as it happens in special relativity. Thus, according to Eqs.(12) and (13), and
taking into account that in the case of photons we can assume $\Lambda = b\;
p$, where $b$ is the impact parameter, we finally obtain
\newpage
\begin{equation} \theta =  \int {dr \over r^2 \sqrt{ 1/b^2 - 1/r^2}}
\end{equation}

Next step is to obtain a simultaneous solution to both Witten's equation (2)
and to Einstein's equation outside matter. For that we assume a static case
with axial symmetry, where $f(x) = f(r)$. Thus, the only non-vanishing
components of the Riemannian connection derived from Eq.(3) are
\begin{eqnarray}\Gamma^0_{01} = \Gamma^0_{10} = \Gamma^1_{00} =
\Gamma^1_{11} = {f'(r)\over 2 f(r)}, \quad\Gamma^1_{22} = -{r^2 f'(r) +
2 r f(r)\over 2 f(r)},\nonumber\end{eqnarray}
\begin{eqnarray} \Gamma^2_{12} = \Gamma^2_{21} = {2 f(r) + r f'(r) \over
2 r f(r)}\end{eqnarray}
where the prime denotes differentiation with respect to $r$. The only nonzero
components of the Ricci tensor are in this case
\begin{eqnarray} R_{00} = { 2 f(r) f'(r) - r [f'(r)]^2 + 2 r f(r) f''(r)
\over 4 r [f(r)]^2},\qquad R_{11} = {[f'(r)]^2 \over [f(r)]^2} - {f'(r)\over
2 r f(r)} - {f''(r)\over f(r)}\end{eqnarray}
\begin{eqnarray} R_{22} = { r^2 [f'(r)]^2 - 4 r f(r) f'(r) - 2 r^2 f(r)
f''(r) \over 4 [f(r)]^2}\nonumber\end{eqnarray}
and the scalar curvature outside matter is
\begin{equation} R = {3 r [f'(r)]^2 - 4 f(r) f'(r) - 4 r f(r) f''(r)\over
2 r [f(r)]^3} \end{equation}
Hence, the only nonzero components of the Einstein tensor outside matter are
\begin{eqnarray} G_{00} = { r [f'(r)]^2 - r f(r) f''(r) - f(r) f'(r) \over
2 r [f(r)]^2}\nonumber\end{eqnarray}
\begin{eqnarray} G_{11} = { 2 f'(r) f(r) + r [f'(r)]^2 \over 4 r [f(r)]^2}
\end{eqnarray}
\begin{eqnarray} G_{22} = { r^2 f(r) f''(r) - r^2 [f'(r)]^2 \over 2 [f(r)]^2}
\nonumber\end{eqnarray}
and the condition $G_{ij} = 0$ leads to the differential equations
\begin{eqnarray}   f'(r) [ r f'(r) - f(r) ] - r f(r) f''(r) = 0, \qquad
f'(r) [ 2 f(r) + r f'(r) ] = 0, \nonumber\end{eqnarray}
\begin{equation} f(r) f''(r) - [f'(r)]^2 = 0\end{equation}
The second equation above has the solutions $f'(r) = 0 $ and $f(r) = A/r^2$,
where $A$ is an arbitrary constant, however, the first and the third equations
above are satisfied if and only if $f'(r) = 0$ {\it i. e.} $f(r) = B = $
constant. Hence the solution to Eqs.(19) is $f = B$.

Otherwise, Einstein's field equation inside matter $R_{ij} - g_{ij} R/2 =
(8\pi\kappa/c^2)\; T_{ij},\;$ ($\kappa$ is Newtons's constant) must be
considered with appropriate values of $T_{ij}$ for a cosmic string in the
interior solution. In this case we assume the metric in the form
\begin{equation} ds^2 = - c^2 dt^2 + a^2 d\theta^2 + a^2 \sin^2\theta d\phi^2
+ dz^2 \end{equation}
where $-\infty \leq t \leq +\infty,\; 0 \leq \theta \leq \theta_M,\; 0 \leq
\phi \leq 2\pi,\; -\infty \leq z \leq +\infty $ and $a$ is the radius of
curvature of a spherical cap inside the string \cite{Got}. For the above
metric the only nonzero connection coefficients are $\Gamma^1_{22} =
- \sin\theta\cos\theta$ and $\Gamma^2_{12} = \Gamma^2_{21} = \cot \theta$ and
the only nonzero components of the Ricci tensor are $R_{11} = 1,\; R_{22} =
\sin^2 \theta$. Thus, the scalar curvature inside the string is
$R = 2/a^2$ and, in consequence of Einstein's equation inside matter, the only
nonzero components of the energy momentum tensor are $T_{00} = - T_{33} =
- c^2/(8\pi\kappa a^2) = - \rho$ where $\rho$ is the volumetric density of
mass. These results lead to a deficit angle produced by the string $\Delta =
2\pi\; ( 1 - \cos\theta_M) = 8 \pi \kappa\mu/c^2$, where $\mu$ is the linear
mass density.

With the exact and unique form $f(x) = B$ before derived, the space-time
metric (3) outside the string becomes
\begin{equation} ds^2 = B\; ( -c^2 dt^2 + d r^2 + r^2 d\theta^2)\end{equation}
however, we can define the new coordinates $ t' = \sqrt{B}\; t $ and $ r' =
\sqrt{B}\; r$, which yield the usual form
\begin{equation} ds^2 =  - c^2 dt^2 + d r^2 + r^2 d\theta^2, \qquad
0 \leq \theta \leq \theta_M\end{equation}
after dropping the primes. Notice that with the space-time metric given in
the above form, and with the assumption $f(x) = B$, the CSG framework before
derived [Eqs.(4) - (12)] is reduced to a free particle approach, as it happens
in GR.

Let us now return to Eq.(14). Its integration yields $ 1/r = b^{-1}\;\cos
\theta$, stating that photons should follow straight lines near a cosmic
string. However, such a solution is not allowed by GR and also by CSG, because
in this latter approach space-time is curved outside the string, whose scalar
curvature depends on $f(x)$. These arguments justify the search of other
possible solutions to Eq.(14). For that we  first transform the integral (14)
into a differential equation, under a change of the integration variable to
$ u(\theta) = 1/r$ :
\begin{equation} \biggl( {du\over d\theta} \biggr)^2 + u^2 = {1\over b^2}
\end{equation}
which yields after a new differentiation
\begin{equation}\biggl({ du\over d\theta }\biggr)\; \biggl({ d^2 u \over
d\theta^2} + u \biggr) = 0 \end{equation}
A first solution to the above equation is $du/d\theta = 0$ {\it i. e.} a
circular path with radius $b$, according to Eq.(23). The solutions to the
remaining equation in Eq.(24) are $ 1/r = b^{-1}\;\cos\theta$ (straight lines,
as before), $ 1/r = b^{-1}\;\sin \theta$ and $ 1/r = b^{-1}\;\exp[i\theta]$.
These latter solutions are meaningless and they must be discarded. The first
interpretation of a circular path is given by GR, which states that photons
follow archs of circles around a cosmic string (open trajectories, since
$0 \leq \theta \leq \theta_M$ defines a wedge). However, the circular path
of photons emerges naturally in CSG, because Eq.(14) does not depend on $f(x)$
although it is derived from the metric (3). Otherwise, since Witten's equation
holds only outside matter, then in the CSG approach we cannot associate the
energy-momentum tensor of the string to a field equation inside matter and
then derive the angular deficit produced by the string, as is done in GR.
Finally, since photons are ``attracted'' by the string in the CSG model and
once we cannot have photons in bound states (closed circular paths), then all
we can assert is that in the CSG framework the path of light near a cosmic
string is also an open circular trajectory.

As a conclusion, we see that in three space-time dimensions, CSG and GR state
that outside matter a first physical solution to the field equation (2) is the
metric (22), which is a trivially conformally transformed of a flat space-time
metric, where $ 0 \leq \theta \leq \theta_M$. The term trivial here means that
$f(x)$ in Eq.(3) is an arbitrary dimensionless constant, when we look for
a result which agrees with GR. However, the form $\ball g_{ij}=f(x)\eta_{ij}$
also yields a circular path for a light ray.

Otherwise, it is important to recall that GR is not a conformally invariant
theory in (3+1)-space-time dimensions \cite{Fro}. This is corroborated by the
present paper, because if the form (3) were extended to four space-time
dimensions, we would conclude that such a metric is not a solution of
Einstein's equation outside matter. Notice that the metric (3), when written
in (3+1)-dimensions, yields a vanishing Weyl tensor (conformally flat
space-time). Moreover, the field equation in the (3+1)-dimensional case is no
longer Eq.(2). It is also important to recall that according to Witten's work
\cite{Wit}, gravitation is a gauge theory in three space-time dimensions,
however this is not true in four space-time dimensions. This suggests that
gauge invariance and conformal transformations should be better investigated
in gravitational theories.

\end{document}